\documentstyle[amssymb,aps,preprint]{revtex}

\begin{document}
\title{Chaos-based cryptograph incorporated with S-box algebraic operation}
\author{Guoning Tang$^{1,2}$, Shihong Wang$^{1,3}$, Huaping L\"{u}$^{1,4},$and Gang
Hu$^{5,1,6\ast }$}
\address{$^1$Department of Physics, Beijing Normal University, Beijing\\
100875,China\\
$^2$Department of Physics, Guangxi Normal University, Guilin 541000,China\\
$^3$Science school, Beijing University of Posts and\\
Telecommunications,Beijing 100876,China\\
$^{4}$Department of Physics, Xuzhou Normal University, Xuzhou 221009, China\\
$^5$China Center for Advanced Science and Technology (CCAST) (World\\
Laboratory), P.O. Box 8730, Beijing 100080, China\\
$^6$The Key Laboratory of Beam Technology and Material Modification of\\
Ministry of Education, Beijing Normal University, Beijing 100875, China\\
$^{*}$Correspondent author (Email: ganghu@bnu.edu.cn)}
\date{\today}
\maketitle

\begin{abstract}
Error function analysis is an effective attack against chaotic cryptograph
[PRE 66, 065202(R) (2002)]. The basin structure of the error function is
crucial for determining the security of chaotic cryptosystems. In the
present paper the basin behavior of the system used in [21] is analyzed in
relation with the estimation of its practical security. A S-box algebraic
operation is included in the chaotic cryptosystem, which considerably
shrinks the basin of the error function and thus greatly enhances the
practical security of the system with a little computational expense.

PACS numbers: 05.45.Vx, 05.45.Ra
\end{abstract}

\section{ Introduction}

In the last decade, secure communication based on chaos synchronization has
attracted continual attention [1-9]. In early stage, scientists in this
field did much effort in inventing methods to hide private message in
chaotic signals and in presenting a variety of designs for chaos
synchronization of encryption and decryption systems. Recently, the study
has gone further into practical aspects. The standard evaluations popularly
used in the conventional secure communication on some essential
cryptographic properties, such as security, performance and robustness
against channel noise and so on, have been applied to chaotic cyptosystems.
With these evaluations some advantages and disadvantages of chaos-based
cryptosystems have been explored [10-18]. In this paper, we will propose a
new chaotic cryptosystem and focus on evaluating its security.

Chaotic cryptograph is based on analytical floating point computation [18],
and it has a property of aperiodicity, which is very much desirable for
secure communication. Though any trajectory of autonomous chaos must be
periodic in computer realizations with finite precision, the actual period
of a system with a number of chaotic coupled oscillators (e.g., ten
oscillators) may be so large that realistic communications can never reach
this period, and thus aperiodicity is practically kept in most of
spatiotemporally chaotic systems [19]. Moreover, chaotic systems have
positive Lyapunov exponents leading to the sensitivity of trajectories to
initial conditions and system parameters, and these features characterize
very good properties of bit diffusion and confusion [17,20]. In particular,
in spatiotemporal chaos with large numbers of positive Lyapunov exponents,
these bit confusion and diffusion are conducted in multiple directions and
high dimensions of variable spaces, and may thus become rather strong. This
is of crucial importance for the high security of cryptosystems. However,
chaotic cryptograph has some serious weakness in its security. First, since
any chaotic output is based on floating-point analytical computation, the
underlying dynamics, including its parameters of secret key, may be pursued
(some times very easily [10-17]). Second, positive Lyapunov exponents
determine certain finite bit confusion rate, this finiteness may not fast
enough to confuse small parameter mismatches and may present certain basin
structures around the secret key (see Fig.1 in [21] and Fig.2 in [22]),
which can be used by intruders for effective attacks. On the other hand,
conventional cryptograph uses algebraic operations on integers, some methods
such as S-box transformation can greatly increase the difficulty of attacks
of analytical computations, and can effectively distort and even eliminate
basin around the secret key. Nevertheless, a single or few S-box operations
cannot make sufficiently strong bit diffusion and confusion, and many rounds
of S-box or other algebraic operations are needed for reaching high security
(e.g., for Advanced Encrytption Standard, AES, one needs 10, 12 and 14
rounds for 128bit, 192bits and 256bit keys, respectively). Therefore, it is
interesting to combine approaches of both chaotic and conventional
cryptographic methods, and to enhance the security of cryptosystems in some
convenient ways. This has been done before by a number of authors[17,18,23].
In this paper, we will apply the idea of S-box in our chaotic cryptograph
which mainly uses floating point analytical computation.

In [21] we suggested to apply spatiotemporal chaos produced by a one-way
coupled one-dimensional map lattice for achieving high practical security.
In [22] a two-dimensional map lattice is used for obtaining efficient
performance, i.e., fast encryption and decryption speeds, by applying
parallel encryption (decryption) operations from different chaotic sites. In
this paper we will go further from [21] by incorporating the S-box \
algebraic operation in the chaotic cryptograph and reach high security with
low computation expense. The paper will be organized as follows. In Sec.II
we specify our system and present some cryptographic results without S-box,
which will be evaluated by the error function. And the basin structure and
its influence on the practical security are analyzed in detail. In Sec.III,
we include a S-box operation, and show that with the combinatorial
applications of spatiotemporal chaos and S-box the key basin of the error
function is effectively shrunk to a single point well in the computation
precision. This leads to great enhancement of the security of the chaotic
crytposystem. In the last section, brief discussion and conclusion of the
practical security of the system against other attacks currently used in
chaotic and conventional cryptanalyses are presented.

\section{ Chaotic cryptosystem and its security evaluation}

We start from the cryptosystem of one-way coupled map lattice used in [21],
with a slight modification by using the nonlinear parameters rather than
coupling parameters as the secret key. The encryption transformation reads

\begin{center}
\begin{eqnarray}
x_{n+1}(j) &=&(1-\varepsilon )f_j[x_n(j)]+\varepsilon f_j[x_n(j-1)],\ \ \  
\eqnum{1a} \\
f_j(x) &=&(3.75+a_j/4)x(1-x),\ \ j=1,...,m  \nonumber
\end{eqnarray}
\begin{eqnarray}
x_{n+1}(j) &=&(1-\varepsilon )f[x_n(j)]+\varepsilon f[x_n(j-1)],  \eqnum{1b}
\\
\ f(x) &=&4x(1-x),\ \ j=m+1,...,N  \nonumber
\end{eqnarray}

\begin{equation}
S_n=(K_n+I_n)\ \ \text{mod }2^\nu ,\ \ K_n=[\text{int}(x_n(N)\times 2^\mu ]\
\ \text{mod }2^\nu   \eqnum{1c}
\end{equation}

\begin{equation}
x_n(0)=S_n/2^\nu   \eqnum{1d}
\end{equation}
\end{center}

and the decryption transformation is

\begin{eqnarray}
y_{n+1}(j) &=&(1-\varepsilon )f_j[y_n(j)]+\varepsilon f_j[y_n(j-1)],\  
\eqnum{2a} \\
f_j(x) &=&(3.75+b_j/4)x(1-x),\ \ j=1,...,m  \nonumber
\end{eqnarray}

\begin{equation}
y_{n+1}(j)=(1-\varepsilon )f[y_n(j)]+\varepsilon f[y_n(j-1)],\ \ \
j=m+1,...,N  \eqnum{2b}
\end{equation}

\begin{equation}
K_n^{^{\prime }}=[\text{int}(y_n(N)\times 2^\mu ]\ \ \text{mod }2^\nu ,\ \ \
I_n^{^{\prime }}=(S_n-K_n^{^{\prime }})\ \ \text{mod }2^\nu   \eqnum{2c}
\end{equation}

\begin{equation}
y_n(0)=S_n/2^\nu =x_n(0)  \eqnum{2d}
\end{equation}
In Eqs.(1) and (2), ${\bf a}=(a_1,a_2,\cdots ,a_m)$ and ${\bf b}%
=(b_1,b_2,\cdots ,b_m)$ serve as the secret keys of encryption and
decryptoin, respectively, which can be freely chosen in the interval $(0,1)$%
. The transmitter sends the ciphertext $S_n$ in the public channel, and the
receiver can achieve chaos synchronization with the transmitter by applying
the decryption key ${\bf b=a}$ and the ciphertext driving $%
y_n(0)=x_n(0)=S_n/2^\nu $, and then correctly receive the plaintext as

\begin{equation}
y_n(N)=x_n(N),\ \ \ K_n^{^{\prime }}=K_n,\ \ \ I_n^{^{\prime }}=I_n 
\eqnum{3}
\end{equation}

We assume that the intruder has the same communication machine, and thus he
knows the structure of Eqs.(1) and (2), and knows also the values of all the
system parameters ($N,m,\mu ,\nu ,\varepsilon $) except the secret key ${\bf %
a}$. Moreover, the intruder has full knowledge (past and present) of
ciphertexts $S_{n}$ because $S_{n}$ is transmitted in the open channel, and
a part of past plaintexts $I_{n}$ of length $T$ (say, from $n=1$ to $n=T$)
due to some chances. With the above messages accessible to the intruder we
have to consider resistance against the public-structure and plaintext-known
attacks. There exist many attacks of this type, among which we will consider
the error function attack (EFA) by computing the following error function
[21]

\begin{equation}
e({\bf b})=\frac 1T\sum\limits_{n=1}^T\left| i_n^{^{\prime }}-i_n\right| ,%
\text{ \ \ \ }i_n=\frac{I_n}{2^\nu },\text{ \ \ \ }i_n^{^{\prime }}=\frac{%
I_n^{^{\prime }}}{2^\nu }  \eqnum{4}
\end{equation}
from which the intruder may try to extract the secret key ${\bf a}$ and then
unmask all future plaintext. Since EFA fully uses all information available
for the legal receiver except the secret key only, it can be regarded as a
very effective attack.

Now we start our numerical simulation with the following parameters fixed
throughout the paper

\begin{equation}
m=4,\ \ \mu =52,\ \ \nu =32,\ \ \varepsilon =0.99,\ \ a_j=0.5,\ \ j=1,2,3,4 
\eqnum{5}
\end{equation}
In Figs.1(a)-(d) we fix $N=25$, $T=10^5$, and $b_2=b_3=b_4=0.5,$ and run the
cryptosystem by varying a single test key parameter $b_1$, and plot $e(b_1)$
vs $b_1$ for different test resolutions. The initial values of both
transmitter and receiver are chosen arbitrarily, and they vary for different
runs. In computing (4) 100 transient iterations are discarded. In Figs.1(c)
and (d) the error function $e(b_1)$ shows a clear basin around the actual
key $b_1=$ $a_1$, and in Figs.1(a) and (b) far away from the basin we have $%
e(b_1)\approx \frac 13$.

Suppose two data sequences $A_{n}$ and $B_{n}$ are completely random, and
uniformly distributed in $[0,1]$ , and they are completely uncorralated from
each other, then the average error between two data sequences is

\begin{eqnarray}
\langle e(A,B)\rangle  &=&\ \lim_{T\rightarrow \infty }\frac 1T%
\sum\limits_{n=1}^T\left| A_n-B_n\right|   \eqnum{6} \\
&=&\frac 13  \nonumber
\end{eqnarray}
The square root of variance of this error for the sequences of length $T$
reads

\begin{equation}
\sigma =\left( \frac 1T\sum\limits_{n=1}^T\left( \left| A_n-B_n\right|
-\langle e\rangle \right) ^2\right) ^{\frac 12}\approx \frac 1{3\sqrt{2T}} 
\eqnum{7}
\end{equation}
It is confirmed that the numerical results of $e(b_1)$ plotted in Fig.1
satisfy both relations Eqs.(6) and (7) in very high precision when $b_1$ is
chosen far from the key basin. This indicates that the output keystream $K_n$
and $K_n^{^{\prime }}$ of Eqs.(1) and (2) have rather good statistical
properties of randomness and uniform distributions in $(0,1)$, and both $K_n$
and $K_n^{^{\prime }}$ are practically independent from each other if $%
\left| b_1-a_1\right| $ is considerably larger than the basin width.

In [21], similar error function basin was observed (Fig.1 in [21]). Now we
will go further to study the functional behavior of this basin with
numerical data, which will be important for predicting the security level of
the system against EFA. We find that $e(b_{1})$ can be presented by a united
empirical formula

\begin{eqnarray}
e(b_1) &=&\frac 13\frac 1{1+(\frac{c_1}{\left| b_1-a_1\right| })^{\alpha
(b_1)}},\ \alpha (b_1)=\frac{1+\frac 53\frac{\left| b_1-a_1\right| }{c_2}}{1+%
\frac{\left| b_1-a_1\right| }{c_2}},\   \eqnum{8} \\
c_1 &=&2.8\times 10^{-12},\ \ \ \ \ \ c_2=4.0\times 10^{-9}  \nonumber
\end{eqnarray}
which is plotted by the solid curves in Figs.2(a) and (b). The formula
prediction of Eq.(8) perfectly coincides with the numerical results of $%
e(b_1)$ [circles, $T=10^3$ for (a) and $10^7$ for (b)] for all ranges from $%
\left| b_1-a_1\right| \ll c_2$ to $\left| b_1-a_1\right| \gg c_2$. In both
limits of large and small $\left| b_1-a_1\right| $ the error function has
interesting scalings of

\begin{eqnarray}
e(b_1) &\propto &\left| b_1-a_1\right| ^{\alpha _1},\ \ \alpha _1=1,\ \
\left| b_1-a_1\right| \ll c_2  \eqnum{9a} \\
\frac 13-e(b_1) &\propto &\left| b_1-a_1\right| ^{-\alpha _2},\ \alpha _2=%
\frac 53,\ \left| b_1-a_1\right| \gg c_2  \eqnum{9b}
\end{eqnarray}

With the basin structure of Figs.1 and 2, the intruder can make EFA as
follows. Suppose the error function has a key basin of width $W$.\ Away from
the key basin, the error function is flat $e(b_{1})\approx \frac{1}{3}$ with
certain fluctuation, the intruder cannot find any tendency toward the
position of the key. Therefore, he can make only brute force attack to find
the key basin with average cost of $\frac{L}{W}$, where $L$ is the space
length available for setting the key (in our case $L=1$). If the intruder
finds the key basin, he can find the tendency in the basin towards the key,
and can easily expose the key position by certain optimal searching methods,
such as adaptive searching. Therefore, the width of the key basin $W$, is an
extremely important quantity for the security of our system against EFA.
This width can be conveniently defined from Eq.(8). According to Eq.(8) the
error function $e(b_{1})$ is monotonously decreasing towards the key.
Nevertheless, as $\left| \frac{1}{3}-e(b_{1})\right| \ll 1$ (i.e., $\left|
b_{1}-a_{1}\right| \gg \frac{1}{3}$) and $T$ is finite, the inevitable
fluctuation of $e(b_{1})$ characterized by Eq.(7) may be considerably larger
than $\frac{1}{3}-e(b_{1})$, and thus can definitely mask the monotonous
tendency of Eq.(8). According to this argument the basin width $W$ may be
defined as being proportional to the distance between the right and left
boundaries $b_{r}$ and $b_{l}$

\bigskip 
\begin{equation}
W=\beta (b_r-b_l)  \eqnum{10a}
\end{equation}

\begin{equation}
\ \frac 13-e(b_{l,r})\approx \frac 1{3\sqrt{2T}}  \eqnum{10b}
\end{equation}
where the multiple factor $\beta $ in Eq.(10a) can be roughly chosen such
that for $\left| b_1-a_1\right| >W$ no any tendency towards the key location
of $e(b_1)$ can be observed. Equations (10) and (9b), together, lead to

\begin{equation}
W\propto T^{\frac 1{2\alpha _2}}\approx T^{0.3}  \eqnum{11}
\end{equation}
For larger $T$ we can certainly have larger width $W$ and need less tests ($%
\frac 1W$) for finding the key basin, but meantime we need more computation
time for each test ($T$ iterations for each test), and need more cost for
collecting longer known plaintext data. Therefore, the total computation
cost for finding the key basin in the space of four parameters ($%
b_1,b_2,b_3,b_4$) reads

\begin{equation}
Cost\approx (\frac 1{W(T)})^4T  \eqnum{12}
\end{equation}

For numerical simulations we define $b_{l\text{ }}$and $b_{r}$ by the first
left and right crossings of $e(b_{1})$ over $\frac{1}{3},$ respectively,
when $b_{1}$ varies from $b_{1}=a_{1}$ to large $\left| b_{1}-a_{1}\right| $%
, and $W$ is defined as $W=10(b_{r}-b_{l})$ according to Eq.(10a). In
Figs.3(a)-(d) we plot $e(b_{1})$ for different $T$'s by keeping other
parameters the same as Fig.1, where $b_{l},$ $b_{r}$ and $W$ are clearly
indicated. The crossing points $b_{l}$ and $b_{r}$ may fluctuate in
different runs, but this fluctuation does not affect the general tendency
between $W$ and $T$. In Fig.3(e) we plot $W$ against $T$ where the solid
line denotes the scaling line

\begin{equation}
\log W=-10.53+0.30\log T,\ \ \ W=2.95\times 10^{-11}T^{0.30}  \eqnum{13}
\end{equation}
while dots respect numerical data. These numerical results fully confirm the
prediction of Eq.(11) which is deduced from the empirical formula Eq.(8) and
the scaling relation Eq.(9b). Inserting Eq.(13) into Eq.(12) we can specify
the cost to break the security of the system. The minimal computation cost
for breaking our system by EFA can be achieved at $T\approx 10^{36}$ and $%
W(T)\approx 1$ that produces

\begin{equation}
Cost\approx 10^{36}\ \text{Iterations}  \eqnum{14}
\end{equation}

\section{Chaotic cryptograph with a S-box operation}

Though the security given by Eq.(14) is rather high (equivalent to the
effective key of 120 bits), the basin of Eq.(8) restricts the security
level. This basin structure is due to the insensitivity of chaos
synchronization to very small mismatch of the corresponding parameters in
analytical computation of chaos.

With smaller system size $N$, the output keystream $K_{n}^{^{\prime }}$ has
even lower sensitivity to the variable of ${\bf b}$ (i.e., has larger basin
width $W$), and accordingly the system has even lower security. In order to
shrink the basin width and further increase the sensitivity of $%
K_{n}^{^{\prime }}$ to small change of the key ${\bf b}$ with a given
smaller system size (i.e., with less computational expenses), one has to
perform some additional (nonanalytical) operation to directly shift the bits
corresponding to small key variation to the position corresponding to larger
key variation. Specifically, we modify Eqs.(1) and (2) as following.

First, we change (1d) and (2d) to

\begin{equation}
x_n(0)=\left\{ 
\begin{array}{c}
S_n/2^\nu ,\ \ \ \ \ \ \ \text{for}\ \ \frac 18\leqslant S_n/2^\nu \leqslant 
\frac 78 \\ 
S_n/2^\nu +\frac 18,\ \ \ \ \ \ \text{for}\ \ S_n/2^\nu <\frac 18 \\ 
S_n/2^\nu -\frac 18,\ \ \ \ \ \ \text{for}\ \ S_n/2^\nu >\frac 78
\end{array}
\right. \ \ \ \   \eqnum{15}
\end{equation}
$\ \ $

\[
y_{n}(0)=x_{n}(0) 
\]
respectively. These changes effectively increase the sensitivity of the
keystreams $K_{n}$ and $K_{n}^{^{\prime }}$ to the secret key ${\bf a}$ and $%
{\bf b}$, respectively, for small (large) and even zero $S_{n}$ (or $%
S_{n}/2^{\nu }=1$). In Eq.(2) such sensitivity is very low for $%
0<S_{n}/2^{\nu }\ll 1$ and $0<1-S_{n}/2^{\nu }\ll 1$ , and this weakness can
be intelligently used by the intruder with the chosen-ciphertext attack
[24]. Second, we add a S-box in the ($m+1$)th map in both transmitter and
receiver, namely, we modify the ($m+1$)th map of Eq.(1) as

\begin{eqnarray}
x_{n+1}^{\prime }(m+1) &=&(1-\varepsilon )f[x_n(m+1)]+\varepsilon f[x_n(m)],
\nonumber \\
Q_n^{^{\prime }} &=&[\text{int}(x_n^{\prime }(m+1)\times 2^\mu ]\ \ \text{%
mod }2^\nu   \eqnum{16a} \\
Q_n &=&Sbox(Q_n^{^{\prime }})  \nonumber
\end{eqnarray}
\begin{equation}
x_{n+1}(m+1)=Q_n/2^\nu   \eqnum{16b}
\end{equation}
where the S-box is defined as follows:

\begin{eqnarray}
A_{1} &=&[(Q_{n}^{^{\prime }}\gg 24)\&255],A_{2}=[(Q_{n}^{^{\prime }}\gg
16)\&255],  \nonumber \\
A_{3} &=&[(Q_{n}^{^{\prime }}\gg 8)\&255],A_{4}=Q_{n}^{^{\prime }}\&255,\  
\eqnum{17} \\
A_{0} &=&A_{1}\oplus A_{2}\oplus A_{3}\oplus A_{4}  \nonumber \\
Q_{n} &=&[A_{0}\ll 24]+[A_{4}\ll 16]+[A_{3}\ll 8]+A_{2}  \nonumber
\end{eqnarray}
In Fig.4 we show the operations of Eq.(17) schematically. The operation $%
x\gg y$ denotes a right shift of $x$ by $y$ bits and the $\&$ operator is
bitwise AND and $\oplus $ means bitwise XOR [25]. After the S-box
transformation of (17) all the 32 bits in $Q_{n}^{^{\prime }}$ take place in
the highest bits in $Q_{n}$ through $A_{0}$. And this operation greatly
enhances the sensitivity of the output $S_{n}$ to the mismatches of the
secret key (these mismatches are naturally reflected in the low bits of $%
Q_{n}^{^{\prime }}$). The decryption transformation (2) has exactly the same
S-box operation as

\begin{eqnarray}
y_{n+1}^{\prime }(m+1) &=&(1-\varepsilon )f[y_n(m+1)]+\varepsilon f[y_n(m)],
\eqnum{18a} \\
\widehat{Q}_n^{^{\prime }} &=&[\text{int}(y_n^{\prime }(m+1)\times 2^\mu ]\
\ \text{mod }2^\nu  \nonumber \\
\widehat{Q}_n &=&Sbox(\widehat{Q}_n^{^{\prime }})  \nonumber
\end{eqnarray}

\begin{equation}
y_{n+1}(m+1)=\widehat{Q}_n/2^\nu   \eqnum{18b}
\end{equation}

The modified chaotic cryptosystem basically performs analytical
floating-point computations on continuous variables. However, its
performance is incorporating with a very simple conventional operation based
on algebraic transformations supported by integer variables. The following
operations in the whole arithmetic of the system Eqs.(1), (2) and (15)-(18)
are of crucial importance for achieving optimal cryptographic properties:

(i) The transmitted signal $S_n$ is defined as integer sequences by the int
operation in Eqs.(1d), (2d), and this operation effectively enhances the
robustness of the communication against noise disturbances in the
transmission channel, and against the round-off errors of both transmitter
and receiver computers. This improvement is extremely important when the
system has very high security (i.e., very high sensitivity to small changes
of system parameters, variables and driving signal).

(ii) We apply modulo operations on some integer variables [Eqs.(1c), (2c),
(16), and (18)]. These operations greatly enhance the sensitivity of chaos
synchronization to the secret key parameters (${\bf a}$ and ${\bf b}${\bf ),}
and considerably increase the difficulty of any attacks based on analytical
computations${\bf .}$

(iii) The S-box operations of Eqs.(16)-(18) dramatically change the basin
structure of Figs.1 and 2, and effectively improve the security of the
chaotic cryptograph. Particular attention will be paid on this point in the
following discussion.

In Figs.5(a)-(d) we do the same as Figs.1(a)-(d), respectively, with
modifications (15)-(18) taken into account. It is remarkable that with the
S-box operation, the continuous basin in Figs.1 and 2 shrinks now to a
singular needle-like structure with zero width. In Fig.5 various refinements
of detection precision up to $2^{-45}$ precision do not help to amplify the
basin width, this is in sharp contrast with Figs.1 where finer detection
precision can show larger and clearer basins [Fig.1(d)].

Another significant point is that the singular needle-like basin structure
cannot be changed by collecting and applying more known past plaintexts,
i.e., by increasing $T$. In Figs.6(a)-(d) we do the same as Figs.3(a)-(d),
respectively, with Eqs.(15)-(18) applied. It is clear that $W=2^{-45}$ is
kept for different $T$'s of which the largest one is up to $T=10^{8}$ ($%
3.2\times 10^{9}$ bits plaintext). For collecting so many data of plaintext
one has to illegally hear a given secure talk for about five days without
any break (in normal voice communication, the transmission speed is about 8k
bits per second). Thus, the practical security of our modified chaotic
cryptosystem against EFA is determined by the key number $2^{180}$ (four key
parameters, each 45 bits) with the system size $N=10$, which is much higher
than that without the S-box operation [$2^{120\text{ }}$given by Eq.(14)]
with considerably larger size $N=25$.

It is emphasized that the S-box of Eq.(17) is very simple and needs little
computation cost. This operation can thus hardly produce effective bit
confusion and diffusion by itself in the conventional cryptograph.
Nevertheless, incorporating with spatiotemporal chaos synchronization this
algebraic transformation can greatly enhance the practical security of
chaotic cryptograph. This is the most significant result of the present
paper that a suitable combinatorial applications of both chaotic and
conventional cryptographies may yield surprisingly high benefit.

\section{Conclusion and discussion}

In conclusion we would like to remark that the modified chaotic system
[Eqs.(1) and (2) with modifications of (15)-(18)] is highly secure against
EFA, which is a very strong attack because with this attack the intruder
effectively uses all information accessible to the legal receiver, except
the secret key only. We assume that the security against EFA may be probably
the actual practical security of the system. Nevertheless, in this
conclusion part we would like to briefly mention the difficulties and the
results (without giving much reasonings) of some other possible attacks in
evaluating our system. The detailed analysis on all these attacks is in
preparation and will be presented elsewhere [27].

First, the keystream $K_{n}$ and, accordingly, the transmitted signal $S_{n}$
have very good statistical properties. They satisfactorily pass the
randomness checking of all the 16 types of NIST Standard evaluations of [26]
for arbitrary plaintexts. The keystream ($K_{n}/2^{\nu }$) has uniform
distributions in the interval (0,1) up to $q$-order ($q$'s from 1 to 4 have
been carefully checked, and even higher orders up to $q=10$ have been
roughly checked). These uniform distributions remain unchanged by widely
changing the secret key and the plaintexts. The probability distribution of
differences between any ciphertexts is practically independent of the
difference of the corresponding plaintexts. Thus, many attacks in
conventional cryptoalaysis such as linear attack and differential attack can
hardly work for our system. And all attacks based on statistics analysis may
not work effectively.

Second, most of attacks used in chaotic cryptography, such as nonlinear
dynamics forecasting and dynamics reconstructs [10-16] are definitely not
efficient for evaluating our system, because these approaches apply
ciphertext--only and private-structure attacks. Without enjoying the
information of system dynamics and known plaintext these attacks are
incomparably weaker. These attacks often reconstructs the system dynamics by
constructing time-delay return maps. Due to the good statistical properties
mentioned above these return maps all have uniform distributions from which
one can hardly find any trace of the system dynamics as well as the secret
key.

Usually, chaotic cryptograph may be easily attacked by analytical
computations in the case of one-round encryption. However, in our case
nonlinearity together with multiplicity of coupled maps makes analytical
attacks difficult. The truncations and S-box operations break simple
analytical relations between the key and the output signals. A simple
analysis shows that attacks based on analytical solution need much more
computational cost than EFA for breaking the security of the system.

Recently, Hu et al proposed a chosen-ciphertext attack to evaluate chaotic
cryptosystems[24], which is very strong. It can be shown that this
chosen-ciphertext attack can break the security of Eqs.(1) and (2) without
S-box operation with much less cost than Eq.(14), i.e., this attack is much
stronger than EFA. However, with the S-box operation and the associated
modifications((15)-(18)), the chosen-ciphertext attack can be made
completely ineffective by including the bits of the secret key (${\bf a}$
for encryption and ${\bf b}$ for decryption). An analysis similar to [24]
shows that the computation cost of the chosen-ciphertext attack of [24] is
again much more expensive than EFA. \ This matter will be discussed in
detail in our future work. It is well known that chosen-ciphertext attacks
are among the strongest attacks if we classify the intensities of various
evaluations[18,24]. Thus, it is reasonable to estimate the practical
security of our system by Eq.(4) and Fig.5.

With respect to the problem of security, our system has a good advantage
over AES. Our system is practically one-time pad cryptograph if its secret
key can be well hidden and the encryption time is much shorter than the
period of computer realization of chaos. On the contrary, AES is definitely
not one-time pad even if its key is kept secret because for AES same inputs
always produce same outputs. Our system has fairly fast encryption speed.
With the C code running a Pentium III (700MHz) computer it can encrypt
50Mbit ciphers (with the key length of 180 bits) which is comparable to that
of AES having a speed of 83Mbit per second with a key of 192 bits by using
the same computer. The encryption speed of our system can be further greatly
enhanced by parallel encryption operations of multiple space units [22].

Finally, we would like to emphasize that our system will not suffer from the
periodicity of finite precision computer realization of chaos. By increasing
the number of coupled maps those period increases rapidly. We certainly
cannot directly detect the period of our system because it is too long. An
estimation, based on small number of coupled maps shows that the period of
our system with $N=10$ and double precision computation is about $10^{70}$
[19]. The currently best computer in the world needs to work more than $%
10^{45}$ years to reach this period, and thus spatiotemporal chaos can be
surely observed in our system for practically unlimited uses.

\newpage References

[1] L. M. Pecora and T. L. Carroll, Phys. Rev. Lett. 64, 821 (1990).

[2] L. M. Cuomo and A. V. Oppenheim, Phys. Rev. Lett. 71, 65 (1993).

[3] L. Kocarev, K. S. Halle, K. Eckert, L. O. Chua and U. Parlitz, Int. J.
Bif. and chaos 2(3), 709 (1992).

[4] L. Kocarev and U. Parlitz, Phys. Rev. Lett. 74(25), 5028 (1995).

[5] J. H.Xiao, G. Hu and Zh. L. Qu, Phys. Rev. Lett. 77, 4162 (1996).

[6] G. Hu, J. H. Xiao and J. Zh. Yang, Phys. Rev. E 56, 2738 (1997).

[7] D. G. Van Wiggeren and R. Roy, Science 279(20), 1198 (1998).

[8] S. Sundar and A. A. Minai, Phys. Rev. Lett. 85(25), 5456 (2000).

[9] J. Garcia-Ojalvo and R. Roy, IEEE trans. Circuits syst. I, 48(12) , 1491
(2001).

[10] K. M. Short, Int. J. Bif and chaos. 4(4), 959 (1994).

[11] G. Perez and H. Cerdeira, Phys. Rev. lett. 74(11), 1970 (1995).

[12] K. M. Short, Int. J. Bif and chaos. 6(2), 367 (1996).

[13] K. M. Short and A. T. Parker, Phys. Rev. E 58, 1159 (1998).

[14] Ch. S. Zhou and C. H. Lai, Phys. Rev. E 60(1), 320 (1999).

[15] Ch.-S. Zhou and C.-H. Lai, Phys. Rew. E 59(6), 6629 (1999).

[16] A. T. Parker, and K. M. Short, IEEE. Trans. Circuits Syst I. 48(5), 624
(2001).

[17] L. Kocarev, IEEE circuits syst magz., 1(3), 6 (2001).

[18] F. Dachselt, and W. Schwarz, IEEE trans. Circuits syst. I, 48(12) ,
1498 (2001).

[19] S.H. Wang, W.R. Lu, H.P. Lu, J. Y. Kuang, and G. Hu (submitted to Phys.
Rev. E)

[20] C. E. Shannon, Bell Syst. Tech. J. 28, 656 (1949).

[21] S.H. Wang, J.Y. Kuang, J.H. Li, Y.L. Luo, H.P. Lu and G. Hu, Phys. Rev.
E 66, 065202 (2002)

[22] S.H. Wang, W.P. Ye, H.P. Lu, J.Y. Kuang, J.H. Li, Y.L. Luo, and G. Hu,
to be published in Commu. Theor. Phys., July (2003), and can be found in
http://arxiv.org, number: nlin.CD/0303026.

[23] R. Matthews, Cryptologia 13, 29 (1989) .

[24] G. J. Hu, Zh. Feng, R. L. Meng, IEEE. Trans. Circuits Syst I. 50(2),
275 (2003).

[25] B. Schneier. Applied Cryptography Second edition: protocols,
algorithms, and source code in C. John Wiley \& Sons, Inc (1996).

[26] A. Rukhin et al, NIST Special Publication 800-22,
http://csrc.nist.gov/encryption/aes.

\strut \lbrack 27] H.P. Lu, et al, in preparation and to be
submitted.\newpage Captions of Figures

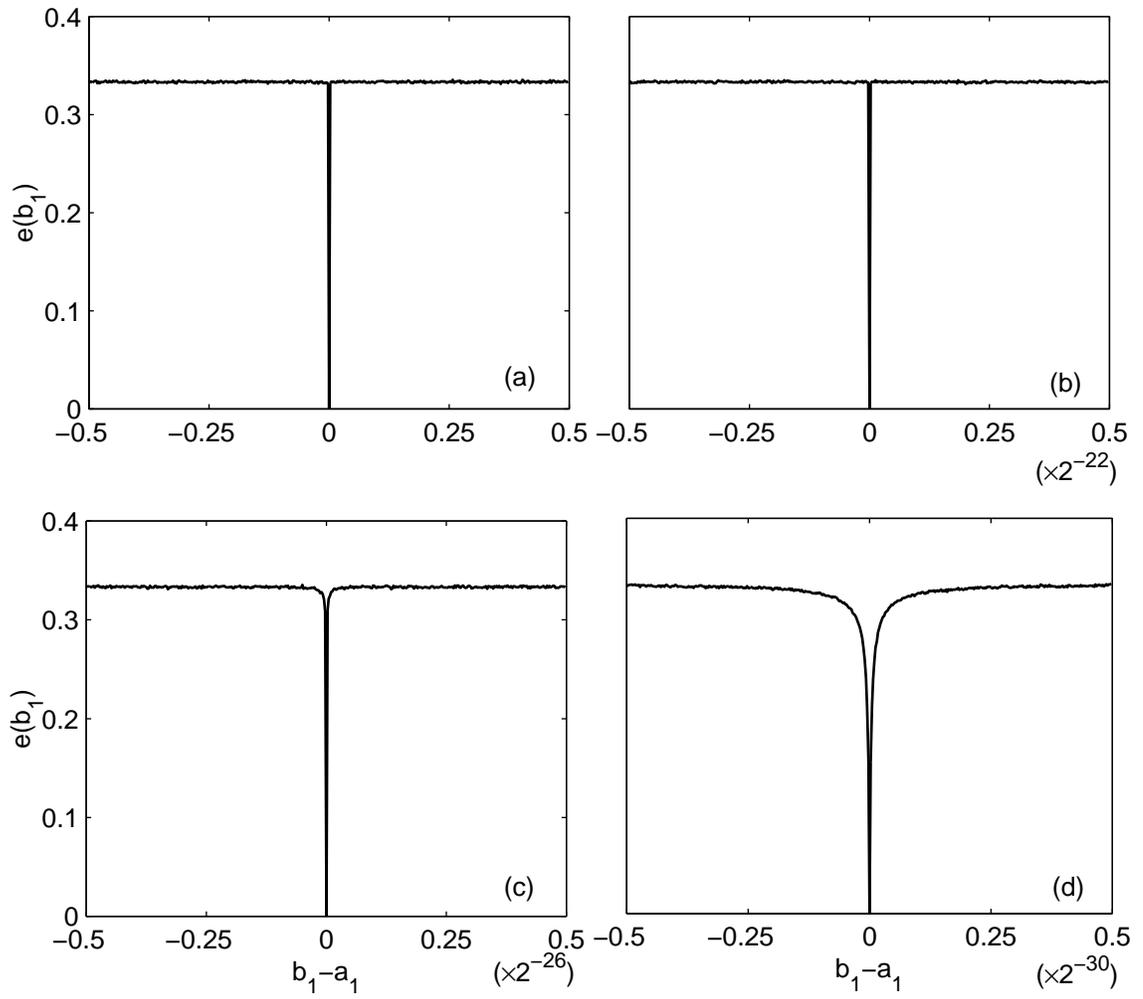
Fig.1. Error function $e(b_1)$ defined in Eq.(4) vs the decryption key
parameter $b_1$ with different detection resolutions. $T=10^5$. The basins
of the error function around the key position $b_1=a_1$ can be clearly
observed in (c) and (d).

Fig.2 (a) and (b) Comparison of the prediction of Eq.(8) (solid lines) with
numerical measurements (circles) of the error function Eq.(4) for $T=10^{3}$
[(a)] and $T=10^{7}$ [(b)], respectively. Agreements of the analytical
predictions with numerical results are satisfactory. The agreement shown in
(b) is particularly desirable since there fluctuation of $e(b_{1})$ is
considerably reduced by taking rather long $T=10^{7}$ ($3.2\times 10^{8}$
bits plaintext).

Fig.3 Widths of the basins of error function $e(b_1)$ for different
available lengths of plaintexts (a) $T=10^3$, (b) $T=10^4$, (c) $T=10^6$,
(d) $T=10^7$, respectively. The boundaries $b_l$ and $b_r$ are defined by
the first crossings of $e(b_1)$ over $e(b_1)=\frac 13$ when $b_1$ varies
from $b_1=a_1=0.5$ to left and right, respectively. And the basin width $W$
is taken as $W=10(b_r-b_l).$ (e) $W=10(b_r$ $-b_l)$ plotted vs $T$. The
solid line is the prediction of Eq.(11) while the circles represent
numerical data. Both coincide with each other satisfactorily.

Fig.4 The schematic figure of the S-box operation of Eq.(17). (a) Four bytes 
$A_{1},A_{2}$, $A_{3}$ and $A_{4}$ divided from $Q_{n}^{\prime }$ of 32
bits, and $A_{0}$ produced from $A_{i}$ via XOR operations. (b) Produce $%
Q_{n}$ of 32 bits from the four bytes $A_{0},A_{4}$, $A_{3}$, and $A_{2}$.

Fig.5 The same as Fig.1 with the modifications (15)-(18) taken into account. 
$T=10^5$. The error function basin shrinks now to a needle-like single
point, irregarding the detection resolution, this is in sharp contrast with
the behavior of Fig.1. In (d) the resolution is up to $2^{-45}$, indicating
that the key number of the four parameters $b_1$ to $b_4$ can be up to $%
2^{180}$.

Fig.6 The same as Figs.3(a)-(d), respectively, with Eqs.(15)-(18) applied.
Increasing $T$ can effectively reduce the fluctuation of $e(b_{1})$, but
cannot change the needle-like basin structure.

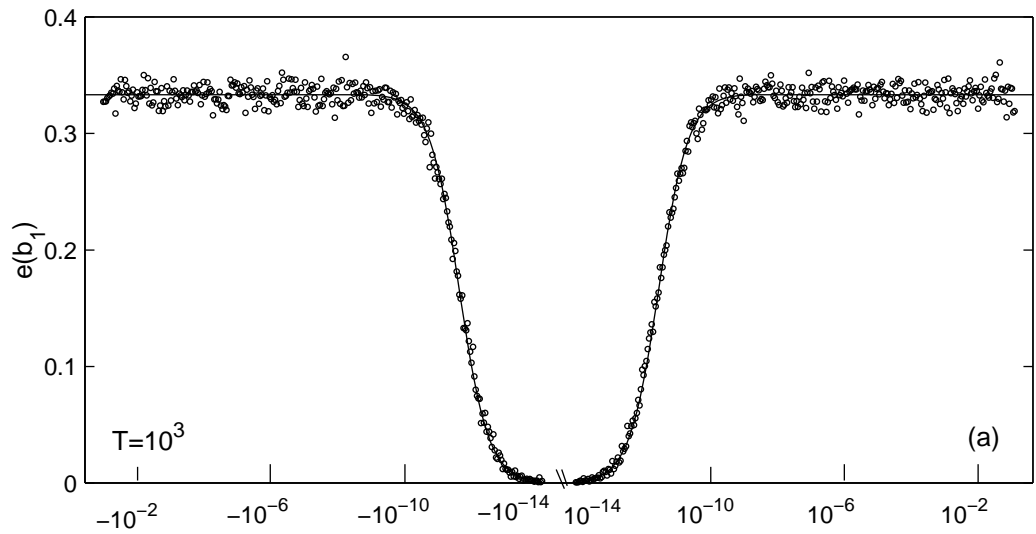
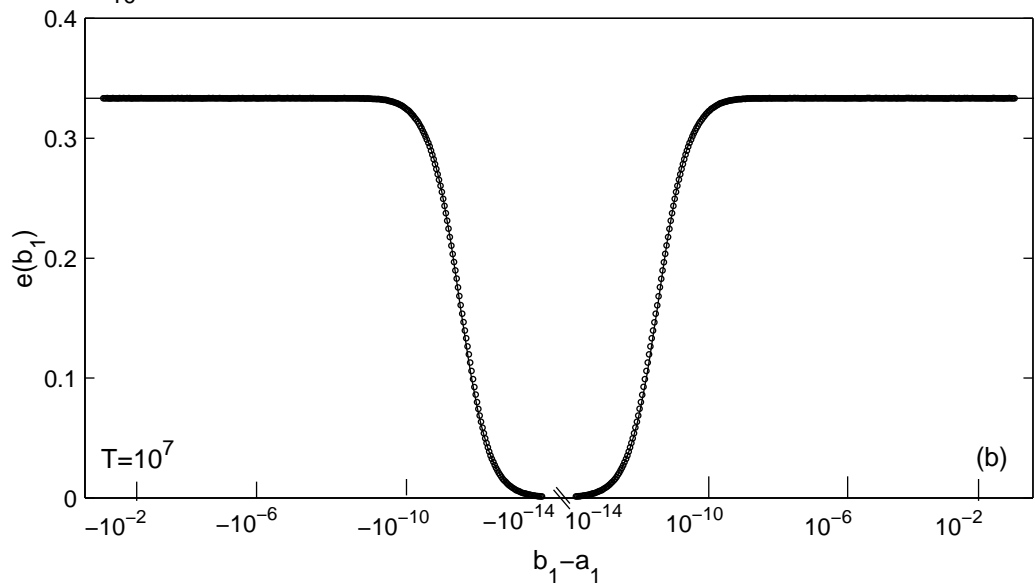
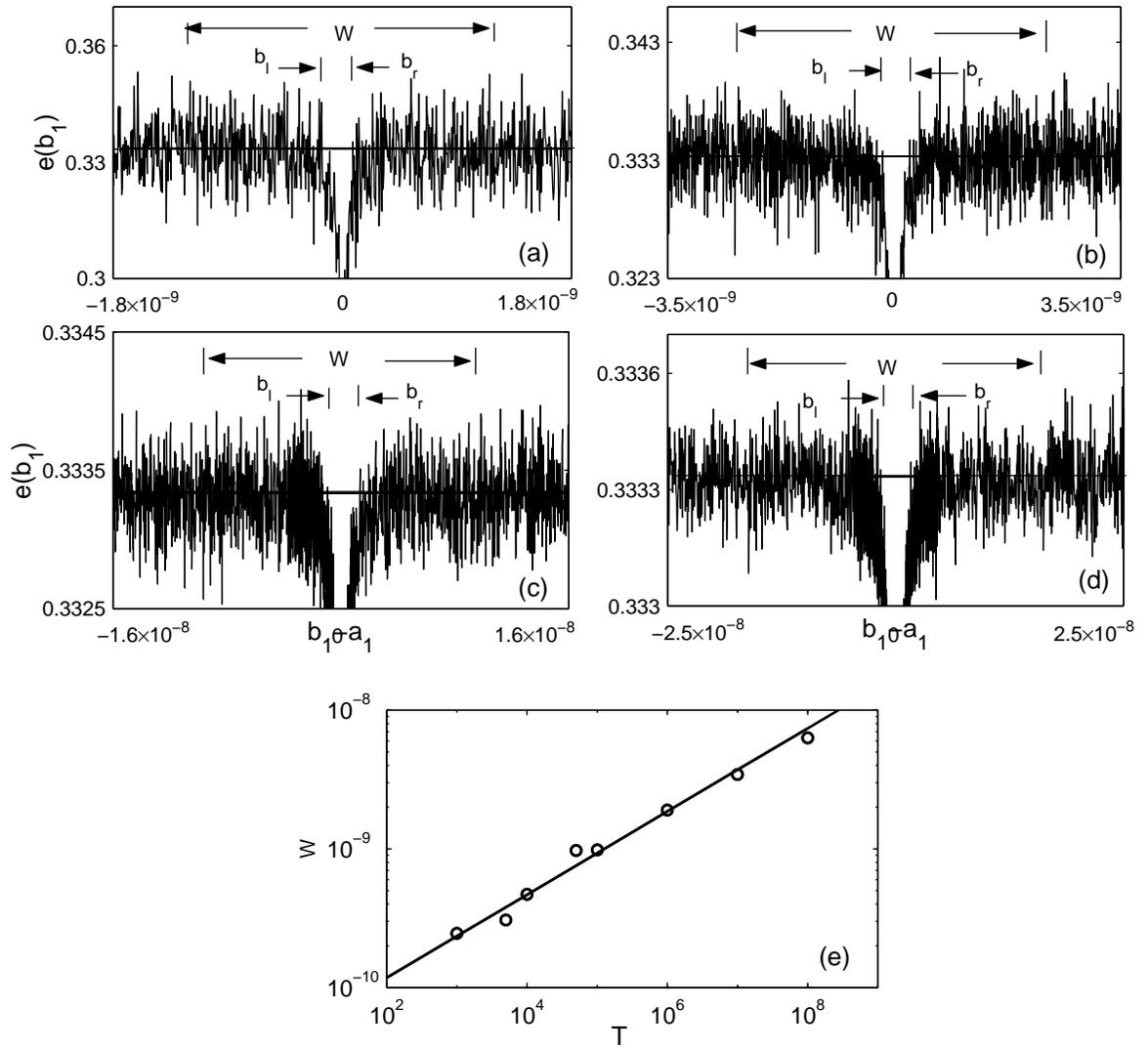
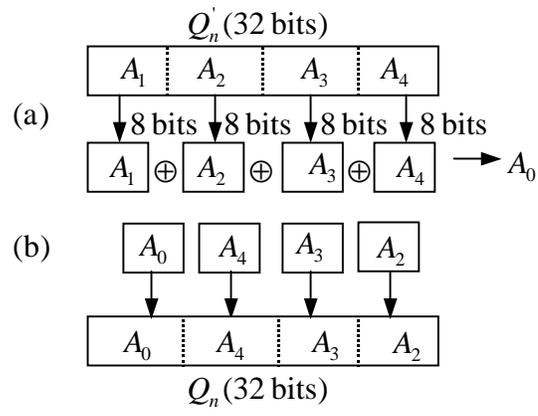
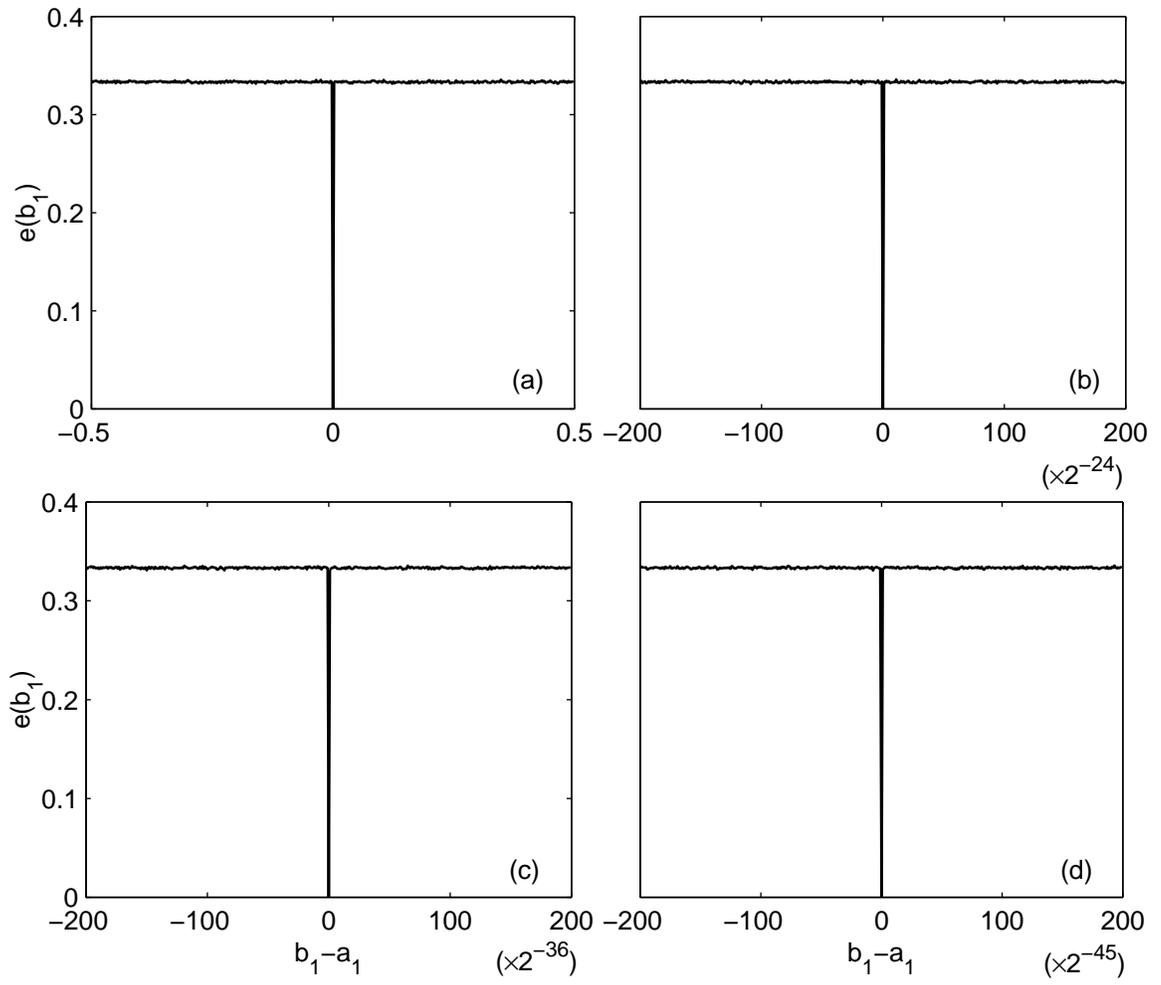
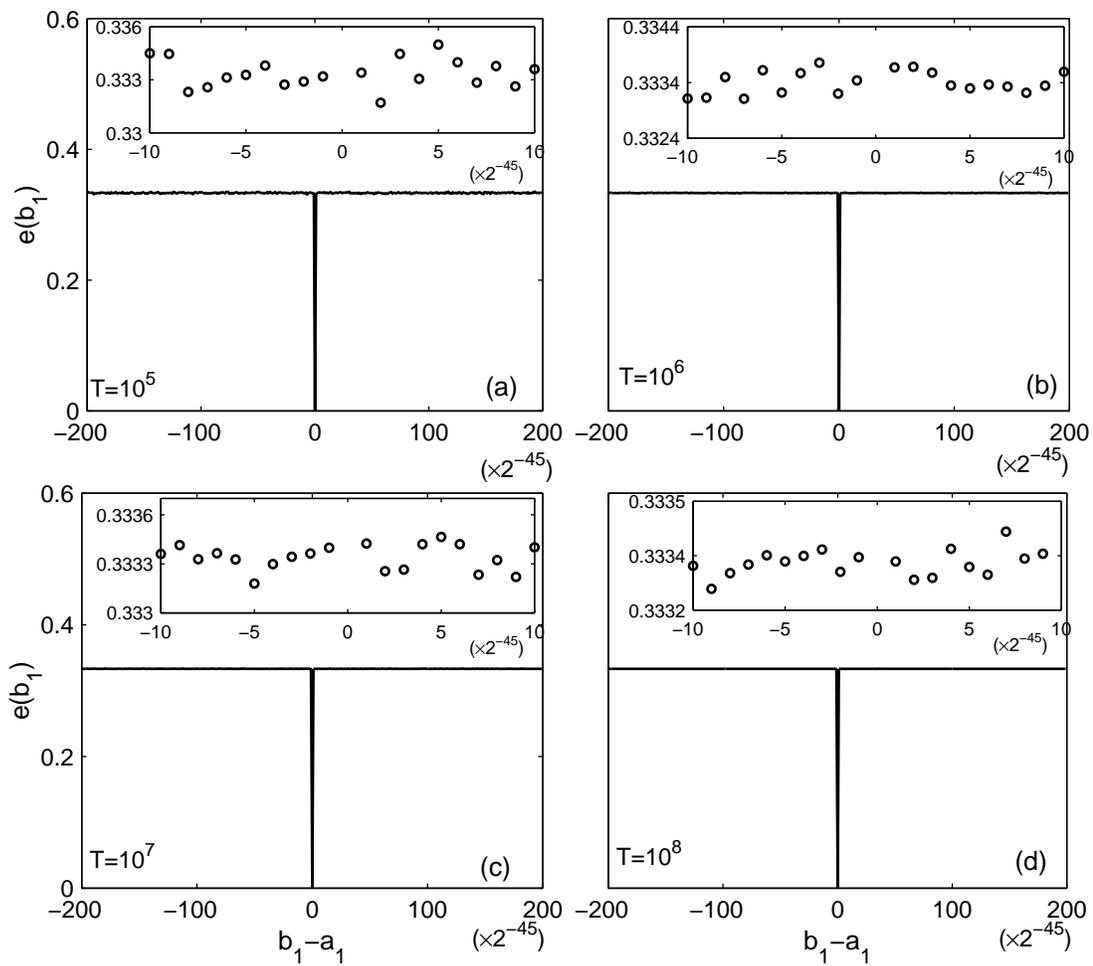
\end{document}